\documentclass[prd,11pt,a4paper,twoside,tightenlines,nofootinbib,showpacs]{revtex4}
\usepackage{latexsym,amsfonts,mathrsfs,amsmath,amssymb,amsopn,bm,mathbbol}

\newcommand{\Lag}{\ensuremath{\mathscr{L}}}
\renewcommand{\d}{\ensuremath{\mathrm{d}}}
\newcommand{\D}{\ensuremath{\mathrm{D}}}
\newcommand{\funcd}[2]{\frac{\delta #1}{\delta #2}}
\newcommand{\ii}{\ensuremath{\mathrm{i}}}
\newcommand{\dyad}[1]{\ensuremath{\overset{\leftrightarrow}{#1}}}

\begin{document}

\author{Hendrik van Hees}
\affiliation{Fakult{\"at} f{\"u}r Physik, Universit{\"a}t Bielefeld,
    Universit{\"a}tsstra{\ss}e, D-33615 Bielefeld}
  
  \title{The renormalizability for massive Abelian gauge field theories
    re-visited} 

\date{May 9, 2003}

\hyphenation{re-pre-sen-ta-tion}

\begin{abstract}
  We give a simplified proof for the perturbative renormalizability of
  theories with massive vector particles. For renormalizability it is
  sufficient that the vector particle is treated as an gauge field,
  corresponding to an Abelian gauge group. Contrary to the non-Abelian case
  one does not need the Higgs mechanism to create the appropriate mass
  terms. The proof uses ``Stueckelberg's trick'' and the Ward-Takahashi
  identities from local Abelian gauge invariance. The simplification is due
  to the fact that, again contrary to the non-Abelian case, no BRST
  analysis is needed.
\end{abstract}
\maketitle

\section{Introduction}

We like to show that the theory of an interacting massive Abelian gauge
field is renormalizable and that this proof can be significantly
simplified, compared to the BRST method, discussed in \cite{rar03}. As we
shall see, we start with an explicitly gauge invariant classical action and
then can use Ward-Takahashi identities (WTIs), directly obtained from this
gauge invariance of the classical action. For the Abelian theory it is not
necessary to use BRST symmetry.

It is also shown that one can choose a gauge, in which both, the
Faddeev-Popov and the Stueckelberg ghost, are free fields and need no
renormalization.

It should be mentioned that these results themselves are not new, since
Kroll, Lee and Zumino \cite{klz67} showed the renormalizability of the
$S$-matrix, making use of the Proca version of the Lagrangian.

Here we show that this analysis of renormalizability can be much simplified
by introducing the Stueckelberg ghost, because then it becomes a gauge
theory which can be treated in manifestly renormalizable gauges. In this
paper we use a gauge, which is a specialization of the well known
$R_{\xi}$ gauges, used for theories with spontaneously broken gauge
symmetries\cite{rxigauge72}.  Then we need only standard methods to prove
the renormalizability of the model making use of the WTIs\cite{wein1}.

\section{The model}

We introduce a scalar field, the \emph{Stueckelberg ghost}
\cite{stueck38a,stueck38b,stueck38c}, into the action of a vector field
with a ``naive mass term'':
\begin{equation}
\label{1}
\Lag_V=-\frac{1}{4} V_{\mu \nu} V^{\mu \nu} + \frac{m^2}{2} V_{\mu} V^{\mu} +
\frac{1}{2} (\partial_{\mu} \phi) (\partial^{\mu} \phi) + m \phi
\partial_{\mu} V^{\mu}.
\end{equation}
Here $V_{\mu}$ is a real vector field and $V_{\mu \nu}=\partial_{\mu}
V_{\nu}-\partial_{\nu} V_{\mu}$ the usual field-strength tensor.

The advantage is that this Lagrangian leads to a manifestly gauge invariant
action
\begin{equation}
\label{2}
S_V[V_{\mu},\phi]=\int \d^4 x \Lag_V(x).
\end{equation}
The local gauge transformation for the fields are given by
\begin{equation}
\label{3}
V_{\mu}' = V_{\mu}+\partial_{\mu} \chi, \quad \phi'= \phi+m \chi,
\end{equation}
where $\chi$ is an \emph{arbitrary} scalar field.

The invariance of the action under infinitesimal transformations is easily
checked:
\begin{equation}
\label{4}
\delta S_V= \delta \int \d^4 x \Lag_V=\int \d^4 x \left [m^2 V_{\mu}
  \partial^{\mu} + m \partial_{\mu} \phi 
\partial^{\mu} + m^2 \partial_{\mu} V^{\mu} + m \phi \partial_{\mu}
\partial^{\mu} \right] \delta \chi=0
 \end{equation}
since the integrand is a total derivative.

It is clear that this gauge transformation remains a symmetry, if ``matter
fields'' are introduced and the vector field is minimally coupled to
conserved currents of these fields. As an example we consider the charged
pions\footnote{To have a specific physical model in mind, one might look at
  the model as a vector-meson dominance model for interacting neutral
  $\rho$-mesons and pions\cite{sak60,klz67,sak72}. It is clear that the
  addition of more Abelian vector fields, as done in these papers, does not
  change the argument. Of course, one has to introduce a Stueckelberg ghost
  for each massive gauge field.} as a complex scalar field in the following
way:
\begin{equation}
\begin{split}
\label{5}
& \Lag_{\pi} = (\D_{\mu} \pi)^*(\D^{\mu} \pi) - m_{\pi}^2 \pi^* \pi -
\frac{\lambda}{8} (\pi^* \pi)^2, \\
& \text{with } \D_{\mu} \pi =(\partial_{\mu} + \ii g V_{\mu} \pi).
\end{split}
\end{equation}
It should be mentioned that the restriction to those minimal couplings of
the vector field to a conserved current is necessary for
renormalizability. While a massless vector field is necessarily a gauge
field, due to the Poincar{\'e} invariance (see, e.g., \cite{wein1}) a
massive vector field does not need to be a gauge field. Nevertheless, as
far as is known today, there exist no renormalizable theories with massive
vector fields, which are not treated as gauge fields.

The pion fields transform under gauge transformations as follows:
\begin{equation}
\label{6}
\pi'=\exp(-\ii g \chi) \pi, \; \pi^*{}'=\exp(+\ii g \chi) \pi^*.
\end{equation}

The four-$\pi$-field interaction term in (\ref{5}) is introduced in order
to keep the quantized theory renormalizable. For our renormalizability
proof, it is important that this is the only superficially renormalizable
self-interaction term for the $\pi$ which is consistent with gauge
invariance.

With the path-integral approach, the quantization is straight forward: In
the following we assume that a gauge invariant regularization, e.g.,
dimensional regularization, is applied. As for any gauge theory, one has to
fix the gauge and introduce Faddeev Popov ghosts. The final expression for
the generating functional of Green's functions is
\begin{equation}
\label{7}
Z[J]=N \int \D \Xi \exp[\ii
  S_{\text{eff}} + J_k \Xi^k].
\end{equation}
Here we introduce $\Xi^k$ as an abbreviation for all the fields, introduced
so far, including the Faddeev-Popov ghosts. 

We choose the following gauge fixing function, inspired by the
$R_{\xi}$-gauges \cite{rxigauge72} used for the treatment of models with
spontaneously broken gauge symmetries, like the standard model. One of its
advantages is the vanishing of the mixing of the Stueckelberg ghost field
with the vector field, introduced into the classical Lagrangian to obtain a
gauge invariant Lagrangian with a massive gauge boson. Another advantage is
that it contains the Proca formulation of the model as the limit $\xi
\rightarrow \infty$ which thus is shown to be just the ``unitary gauge''
for our Abelian gauge theory.
\begin{equation}
\label{8}
g=\partial_{\mu} V^{\mu} + \xi m \phi.
\end{equation}
In this gauge the effective Lagrangian is given by
\begin{equation}
\begin{split}
\label{9}
\Lag_{\text{eff}} = & \Lag_{\pi} -\frac{1}{4} V_{\mu \nu} V^{\mu \nu} +
\frac{m^2}{2} V_{\mu} V^{\mu} - \frac{1}{2 \xi} (\partial_{\mu}
V^{\mu})^2 \\
& + \frac{1}{2} (\partial_{\mu} \phi)(\partial^{\mu} \phi) - \frac{\xi
  m^2}{2} \phi^2 + (\partial_{\mu} \eta)^*(\partial_{\mu} \eta) - \xi m^2
\eta^* \eta.
\end{split}
\end{equation}
The fields $\eta$ and $\eta^*$ are Grassmann fields (Faddeev-Popov ghosts),
and the $J_k$ are external currents. 

The path integral can be read as a usual path integral for vacuum quantum
field theory or at finite temperature. In the latter case one the time
variable has to be read as an imaginary quantity, running from $0$ to $-\ii
\beta$. Thereby the path integral is taken over all field configurations 
with the Kubo-Martin-Schwinger (KMS) conditions:
\begin{equation}
\begin{split}
\label{kms}
\pi(-\ii \beta,\vec{x}) &= \exp(- \beta \mu) \pi(0,\vec{x}), \quad \pi^*(-\ii
\beta,\vec{x})=\exp(\beta \mu) \pi^*(0,\vec{x}), \\ \phi(-\ii
\beta,\vec{x}) &= \phi(0,\vec{x}), \quad V_{\mu}(-\ii \beta,\vec{x})=V_{\mu}(0,
\vec{x}).
\end{split}
\end{equation}
Here $\mu$ is a chemical potential for the conserved charge of the pions.
 
Thereby, it is important to note that the KMS condition for the
Faddeev-Popov ghosts is periodic rather than anti-periodic, because they
are not introduced by the path integral treatment of physical fermions
(see, e.g., \cite{wein1}), but to treat the appearance of the Faddeev-Popov
determinant in the path-integral measure as a contribution to the effective
classical action for a fixed gauge:
\begin{equation}
\label{kms-fp}
\eta(-\ii \beta,\vec{x})=\eta(0,\vec{x}), \quad \eta^*(-\ii
\beta,\vec{x})=\eta^*(0,\vec{x}).
\end{equation}
Concerning renormalization, it is sufficient to look at the vacuum case,
since the renormalization of the model is completely done at $T=0$. Going
to finite temperature does not introduce new UV divergences (see, e.g.,
\cite{kap89,lebel}).

The Lagrangian (\ref{9}) is superficially renormalizable since it contains
only terms of momentum power $4$ and less. So we know that the
counterterms, necessary to render the quantum action finite, are local and
also of momentum power $4$ and less, since thanks to our choice of gauge
the propagator of the gauge field goes like $1/p^2$ for large momenta.

Of course, one has to show that the model is really renormalizable since
the classical action is restricted to gauge invariant couplings and a
superficially necessary counterterm like, e.g., $(V_{\mu} V^{\mu})^2$ is
not needed to renormalize the effective action. On the other hand (\ref{9})
is the most general superficially renormalizable Lagrangian, which is
consistent with the local gauge invariance and the field content taken
into account.

To show the renormalizability one can use the BRST invariance as described
in \cite{rar03}. This has the advantage that it is also applicable to
non-Abelian gauge fields. For our purposes it is more convenient to use
directly the local gauge invariance of the classical action. The reason is
that both, the scalar auxiliary (or Stueckelberg) field and the
Faddeev-Popov ghosts, are free fields in (\ref{9}). Thus, we can define the
gauge transformation to act trivially on the ghost fields, i.e., leaving
them unchanged. As we shall show in a moment, this prevents the WTIs from
gauge invariance for the generating functional $Z$ to be free of higher
order derivatives with respect to the external currents $J_k$. It is clear
that for non-Abelian theories this holds no longer true, since the
Faddeev-Popov ghosts, at least, are coupled to the gauge fields. Then the
analysis through the BRST formalism is more appropriate for the
renormalizability proof (see, e.g., \cite{wein2}).

\section{The Ward-Takahashi identities}

As shown in the previous section, the effective classical action (\ref{9}),
appearing in the path integral (\ref{7}), is given by the classical gauge
invariant Lagrangian, cf. (\ref{1}) and (\ref{5}), the gauge-fixing, and
the source part:
\begin{equation}
\begin{split}
\label{10}
\Lag_{\text{fix}} &= -\frac{1}{2 \xi} (\partial_{\mu} V^{\mu} + \xi m
\phi)^2, \\
\Lag_{\text{src}} &= J_k \Xi^k := j_{\mu} V^{\mu} + k \phi + \eta^* \alpha
+ \alpha^* \eta + a^* \pi + \pi^* a.
\end{split}
\end{equation} 
For completeness we note also the Lagrangian for the free Stueckelberg
and the Faddeev-Popov fields:
\begin{equation}
\label{11}
\Lag_{\text{FP$\phi$}}=\frac{1}{2} (\partial_{\mu} \phi)(\partial^{\mu}
\phi) - \frac{\xi m^2}{2} \phi^2 + (\partial_{\mu} \eta^*)(\partial_{\mu}
\eta) - \xi m^2 (\eta^* \eta).
\end{equation}
In this notation the path integral (\ref{9}) reads
\begin{equation}
\label{12}
Z[J]=N \int \D \Xi \exp[\ii (S_{V} + S_{\pi} + S_{\text{fix}} + S_{\text{src}} +
S_{\text{FP$\phi$}})].
\end{equation}
The integral over the Faddeev-Popov and the Stueckelberg fields yields just
the generating functional for free fields. This is only important for the
partition sum at finite temperature, since it cancels contributions from
the unphysical degrees of freedom of the vector field, leading to the
partition sum of three bosonic field degrees of freedom with mass $m$, as
it should be\footnote{Of course, for the model of free massive vector
  fields, this can be derived directly by using the Proca Lagrangian
  without the gauge theoretical treatment.}.

For the proof of renormalizability we use the invariance of the
path-integral measure under the gauge transformations, given by the Eqs.
(\ref{3}) and (\ref{6}). By definition the Faddeev-Popov ghosts are
unchanged under the here considered gauge transformations. Then only
$S_{\text{fix}} + S_{\text{src}}$ is gauge dependent. Thus, substituting
gauge transformed fields into the path integral (\ref{12}), for
infinitesimal gauge transformations one obtains the Ward-Takahashi
identities (WTIs) which are summarized by the one identity for the
generating functional $Z$:
\begin{equation}
\label{13}
-\frac{\Box + \xi m^2}{\xi} \left [ \partial^{\mu} \frac{\delta Z}{\ii
 \delta j^{\mu}} + \xi m \frac{\delta Z}{\ii \delta k} \right ]- \partial_{\mu}
 j^{\mu} Z + k m Z -\ii g \left [a^* \frac{\delta Z}{\ii \delta a^*} - a
 \frac{\delta Z}{\ii \delta a} \right ]=0. 
\end{equation}
The next step is to introduce the generating functional for connected
Green's functions $W=-\ii \ln Z$. From (\ref{13}) one immediately obtains
\begin{equation}
\label{14}
-\frac{\Box + \xi m^2}{\xi} \left [\partial^{\mu} \frac{\delta W}{
 \delta j^{\mu}} + \xi m \frac{\delta W}{\delta k} \right ] - \partial_{\mu}
 j^{\mu} + k m -\ii g \left [a^* \frac{\delta W}{\delta a^*} - a
 \frac{\delta W}{\delta a} \right ]=0. 
\end{equation}
Finally, we define the generating functional for one-particle irreducible
(1PI) truncated Green's functions (proper vertex functions) by a functional
Legendre transform:
\begin{equation}
\begin{split}
\label{15}
& \Gamma[\bar{\Xi}] = W[J] - \int \d^4 x J_k(x) \bar{\Xi}^{k}(x), \\
& \bar{\Xi}^k = \funcd{W}{J_k} \; \Leftrightarrow \;
\funcd{\Gamma}{\bar{\Xi}^k} = -J_k.
\end{split}
\end{equation}
From (\ref{14}) we obtain the WTIs for $\Gamma$
\begin{equation}
\label{16}
-\frac{\Box+\xi m^2}{\xi} (\partial_{\mu} \bar{V}^{\mu}+\xi m \bar{\phi}) +
 \partial^{\mu} \funcd{\Gamma}{\bar{V}^{\mu}} - m
 \funcd{\Gamma}{\bar{V}^{\mu}} -m \funcd{\Gamma}{\bar{\phi}} + \ii g \left
 [\funcd{\Gamma}{\bar{\pi}} \bar{\pi} - \funcd{\Gamma}{\bar{\pi}^*}
 \bar{\pi}^* \right ]=0.
\end{equation}

\section{Proof of renormalizability}

To prove the renormalizability of the model we use (\ref{16}) to show that
it is sufficient to introduce a counterterm Lagrangian of the same form as
(\ref{9}), except that we do not need counterterms for the Faddeev-Popov
and Stueckelberg ghosts, since these are free fields and thus not involved
in divergent loop integrals:
\begin{equation}
\begin{split}
\label{17}
\Lag_{\text{ct}}= & -\frac{1}{4} \delta Z_3 V_{\mu \nu} V^{\mu \nu} + \frac{\delta
  m^2}{2} V_{\mu} V^{\mu}  - \delta(1/\xi) (\partial_{\mu} V^{\mu})^2 \\
& + \delta Z_2 (\partial_{\mu} \pi^*) (\partial^{\mu} \pi) - \frac{\delta
  m_{\pi}^2}{2} \pi^* \pi \\
& + \ii g \delta Z_1 [\pi^* \dyad{\partial}_{\mu} \pi] V^{\mu} + g^2 \delta
Z_1' V_{\mu} V^{\mu} \pi^* \pi - \frac{\delta \lambda}{8} (\pi^* \pi)^2.
\end{split}
\end{equation}
We have to show that these counterterms are sufficient to render all
divergences finite. Additionally, due to the gauge invariance of the
classical action $S_V+S_{\pi}$, the counterterms have to satisfy the
\emph{Ward identities} 
\begin{equation}
\label{17b}
\delta Z_1=\delta Z_2=\delta Z_1'
\end{equation}
For sake of simplicity, we consider only the case of unbroken gauge
symmetry. It is no problem to generalize the proof for the case of
spontaneous symmetry breaking, i.e., the massive model with a Higgs
field\footnote{The most simple realization would be to set $m_{\pi}^2<0$,
  analogous to the Higgs fields in the minimal standard model. Then one
  should impose a mass-independent renormalization scheme \cite{kug77}. The
  possibility of such a choice proves that one can subtract all divergences
  in the symmetric phase, thereby introducing a renormalization $\pi$-mass
  scale.}.

The proof is by induction in the loop order $L$. At tree level, $L=0$, all
diagrams are finite. Now we suppose that (\ref{17}), with the restrictions
by the Ward identities (\ref{17b}), is sufficient to render the diagrams up
to loop order $L$ finite. We have to show that then the same holds true
also at loop order $L+1$. Then the renormalized effective action up to loop
order $L$ fulfills the WTI (\ref{16}), including the counterterms up to
this order.

By assumption, for any diagram of loop order $L+1$ we can subtract the
\emph{proper subdivergences}, which come from proper subdiagrams with at
most $L$ loops, with counterterms of the structure (\ref{17},\ref{17b}).
After this subtraction, due to Weinberg's theorem \cite{wein60}, the only
divergences left are the overall divergences, which can appear only for
diagrams with at most $4$ external legs. This means that, in momentum
representation, the divergent parts are polynomials in the external
momenta. The counterterms for $\Gamma$ are local and polynomials in the
fields up to order $4$. We have to prove that these polynomials, for the
contributions to $\Gamma$ at loop order $L+1$, are of the form (\ref{17})
and fulfilling the Ward identities (\ref{17b}). 

We start the proof with the remark that there is no $\phi V$-mixing in
$\Gamma$, i.e.,
\begin{equation}
\label{18}
\left. \frac{\delta^2 \Gamma}{\delta V^{\mu} \delta \phi} \right
|_{\bar{\Xi}=0} = 0.
\end{equation}
Indeed, thanks to our choice of the gauge fixing functional (\ref{8}),
there is no such term at tree level, and the field $\phi$ is free, so that
there are no proper vertex functions involving $\phi$. Thus, all
proper vertex functions vanish except the two-point vertex, which is the
inverse full propagator for the Stueckelberg ghost, which has to be
identical to the free one to all loop orders.

Indeed, taking the functional derivative of the WTI (\ref{16}) with respect
to $\phi$, one finds after setting the mean fields to $0$ and making use of
(\ref{18}):
\begin{equation}
\label{19}
G_{\phi}^{-1}(x,y)=\left . \frac{\delta^2 \Gamma}{\delta \phi(x) \delta
  \phi(y)} \right |_{\bar{\Xi}=0} = -(\Box_x+\xi m^2) \delta^{(4)}(x-y).
\end{equation}
This means that $G_{\phi}$ is indeed the free propagator for a scalar field
with mass $\sqrt{\xi} m$. Thus (\ref{18}) is consistent with the underlying
gauge invariance of the classical action: In this gauge, the Stueckelberg
field needs no renormalization at all, neither the wave function nor the
mass.

The WTI for the vector-boson propagator is obtained by taking the
derivative of the WTI (\ref{16}) with respect to the vector field:
\begin{equation}
\label{20}
\partial_{x \mu} (G_V^{-1})^{\mu \nu}(x,y) = \frac{\Box + \xi m^2}{\xi}
\partial_x^{\nu} \delta^{(4)}(x-y).
\end{equation}
Introducing the polarization (self-energy) of the vector field by
$G_V^{-1}=\Delta_V^{-1} - \Sigma_V^{-1}$, where $\Delta_V$ is its free propagator, and
Fourier transforming (\ref{20}) gives the transversality of the vector
self-energy:
\begin{equation}
\label{21}
p_{\mu} \Sigma_V^{\mu \nu}(p) = 0.
\end{equation}
Thus, at loop order $L+1$ the term, quadratic in $V_{\mu}$, needs a
counterterm of the form
\begin{equation}
\label{22a}
\frac{\delta Z_3}{2} V^{\mu} (\partial_{\mu} \partial_{\nu} - \Box g_{\mu
  \nu}) V_{\nu}.
\end{equation}
In other words, to render the $V$ self-energy finite, the subtraction of a
wave-function renormalization term is sufficient. Thus, the self-energy can
be written in the form
\begin{equation}
\label{22}
\Sigma_V^{\mu \nu}(p) = \left (-p^2 g^{\mu \nu} + p^{\mu} p^{\nu}
\right ) \Sigma_V(p),
\end{equation}
where $\Sigma_V$ is a scalar function of $p$ which is only logarithmically
divergent\footnote{Of course, here also Lorentz invariance was used. At
  finite temperature, the Lorentz invariance is broken by the existence of
  the rest frame of the heat bath. Then we have two scalar functions for
  the polarization tensor, one which is transverse and one which is
  longitudinal with respect to the three-momentum. This does not affect the
  form (\ref{22a}) of the counter term since, after subtraction of the
  subdivergences of a diagram, the remaining overall divergent parts are
  independent of temperature and thus take the corresponding vacuum
  values.}. Thus, one neither needs a vector mass nor a gauge fixing
constant renormalization:
\begin{equation}
\label{23}
\delta m^2=0, \quad \delta (1/\xi)=0.
\end{equation} 
For the $\pi$-propagator, there is no restriction by gauge invariance.
Since it is quadratically divergent, the counter-terms in (\ref{16}) are
sufficient and consistent with the requirements of gauge invariance.

Now we have to consider the vertices. First, we look for those types which
are already present at tree level: Taking the derivative of the WTI
(\ref{16}) with respect to $\bar{\pi}^*(y)$ and $\bar{\pi}(z)$ we obtain
\begin{equation}
\label{24}
\partial_x^{\mu} \Gamma_{\mu}^{V,\pi^*,\pi}(x,y,z)=g G_{\pi}^{-1}(y,z) [
  \delta^{(4)}(x-z) - \delta^{(4)}(x-y) ].
\end{equation}
Here we have used the general definition for a proper vertex function:
\begin{equation}
\label{25}
\Gamma^{\Xi_1,\ldots,\Xi_k}(x_1,\ldots,x_k) = \left . \ii \frac{\delta^k
  \Gamma}{\delta \bar{\Xi}_1(x_1) \cdots \delta \bar{\Xi}_k(x_k)} \right
  |_{\bar{\Xi}=0}.
\end{equation}
Fourier transformation of (\ref{24}) yields
\begin{equation}
\label{26}
\ii p^{\mu} \Gamma_{\mu}^{V,\pi^*,\pi}(p,q,r)= g [G_{\pi}^{-1}(q)-G_{\pi}^{-1}(r)].
\end{equation}
Since the superficial degrees of divergence for the involved vertex
functions are $\delta(\Gamma^{V,\pi^*,\pi})=1$ and $\delta(G_{\pi}^{-1}) =2$,
the overall divergences for these quantities are of the form
\begin{equation}
\label{27}
\Gamma^{V,\pi^*,\pi}_{\text{div}}(p,q,r) = -\ii (C_1 q^{\mu} + C_2 r^{\mu}), \quad
(G_{\pi}^{-1})_{\text{div}}(p)=\delta Z_2 p^2 - \delta m_{\pi}^2.
\end{equation}
For the vertex, here we used $q$ and $r$ as the independent momenta\footnote{Of
  course overall momentum conservation $p+q+r=0$, with all currents running
  into the vertex, is implied.}. 
Applying the WTI (\ref{24}) to the diverging part gives
\begin{equation}
\label{28}
(q_{\mu} - r_{\mu})(C_1 q^{\mu}+C_2 r^{\mu}) = \delta Z_2 g(q^2-r^2).
\end{equation}
Comparing both sides of this equation yields
\begin{equation}
\label{29}
C_1=C_2:=\delta Z_1 g = \delta Z_2 g \; \Rightarrow \; \delta Z_1=\delta Z_2.
\end{equation}
Thus the first of the Ward identities (\ref{17b}) is fulfilled also at loop
order $L+1$.

Derivation of (\ref{16}) with respect to $\pi^*(y)$, $\pi(z)$, and
$A^{\nu}(x')$ gives, after a Fourier transformation, the WTI
\begin{equation}
\label{30}
p^{\mu} \Gamma_{\mu \nu}^{V,V,\pi^*,\pi} (p,p',q,r) =
\Gamma_{\nu}^{V,\pi^*,\pi}(p,p-q,r) - \Gamma_{\nu}^{V,\pi^*,\pi}(p',q,p+r).
\end{equation}
Since this vertex is logarithmically divergent, the overall diverging part
at loop order $L+1$ is of the form $2 \ii \delta Z_1' g^{\mu \nu}$. Together
with (\ref{27}) this proves the second of the Ward identities (\ref{17b}):
\begin{equation}
\label{31}
\delta Z_1=\delta Z_1'.
\end{equation}
To show that the four-pion interaction counterterm has to be of the given
form, even global gauge invariance is sufficient.

Now we have to show that the superficially divergent vertices, which are
not contained in the original Lagrangian, are finite. First, we see that
$VVV$-vertices are vanishing loop order by loop order, because the
Lagrangian is invariant under charge-conjugation transformations. This is
the same argument as used to prove Furry's theorem for QED, which says that
all vertices with an odd number of vector-field legs (and no other external
legs) vanish.

The symmetry under global gauge transformations also excludes $\pi V V$-
and $\pi\pi\pi V$-vertices. Thus, we only need to show that the vertex with
four $V$-legs is finite (in QED this is the famous Delbr{\"u}ck-scattering
term), although it looks logarithmically divergent. For that we take the
derivative of the WTI (\ref{16}) with respect to $V_{\nu}(w)$,
$V_{\rho}(y)$, and $V_{\sigma}(z)$, which leads, after taking the Fourier
transform, to
\begin{equation}
\label{32}
p^{\mu} \Gamma_{\mu \nu \rho \sigma}^{VVVV}(p,q,r,s)=0.
\end{equation}
Since the vertex is logarithmically divergent, the overall divergent part
at loop order $(L+1)$ must be of the form
\begin{equation}
\label{33}
D(g_{\mu \nu} g_{\rho \sigma} + g_{\mu \rho} g_{\nu \sigma} + g_{\mu
  \sigma} g_{\nu \rho}).
\end{equation}
Applying the WTI (\ref{32}) to the divergent part shows $D=0$. This
completes our proof of perturbative renormalizability.

As usual, the gauge independence of the $S$-matrix can be shown by making
use of LSZ reduction, leading to the the so called \emph{equivalence
  theorem} (see, e.g., the article by B. W. Lee in \cite{houches75}). In
our context this means that the renormalized $S$-matrix is independent of
$\xi$ and thus, after renormalization, remains finite for $\xi \rightarrow
\infty$, which leads back to the theory in the Proca form.  Thus, the
$S$-matrix is renormalizable when calculated with the Proca form of the
model which corresponds to the ``unitary gauge'' in Higgs
models\footnote{For a detailed discussion of renormalization in the unitary
  gauge see \cite{koeg93}, where the Stueckelberg formalism is used to
  reconstruct the gauge invariant form of the standard model Lagrangian in
  unitary gauge.}.  The
same time, this shows that our gauge model indeed describes a massive
vector boson, interacting with matter particles (here described by the
$\pi$-fields). In the limit of zero vector-boson mass one obtains Quantum
electrodynamics (in our case ``scalar electrodynamics'').  Of course,
taking the limit $m \rightarrow 0$, one has to take care of possible
infrared problems. It is clear that the same analysis goes through with
other matter contents like Dirac fields.

The gauge independence of observable thermodynamic quantities is seen
simply by the fact that they are thermal expectation values of gauge
invariant operators which can be calculated with help of the path integral
for vanishing external sources. By construction, this path integral is
independent of the chosen gauge fixing condition $g[A,\phi]=0$.

\section{Conclusion}

It was shown that the renormalizability proof for massive Abelian gauge
theories can be significantly simplified by using the Stueckelberg-field
approach.

An extension of this proof to non-perturbative treatments, based on the 2PI
formalism, which generalizes the analysis for theories with global
symmetries \cite{vHK2001-Ren-III}, is in preparation for a later
publication.

\begin{flushleft}

\end{flushleft}

\end{document}